# Effects of receptor clustering on ligand dissociation kinetics: Theory and simulations

**Manoj Gopalakrishnan** [(1,2)], **Kimberly Forsten-Williams**[4] [(3)], **Matthew A. Nugent** [(4)] and **Uwe C. Täuber** [(2)]

1 Department of Biological Physics[5],
   Max-Planck-Institut für Physik komplexer Systeme,
   Nöthnitzer Straβe 38,
   01187 Dresden, Germany.

2 Department of Physics and Center for Stochastic Processes in Science and Engineering,
   Virginia Polytechnic Institute and State University,
   Blacksburg, VA 24061, USA.

3 Department of Chemical Engineering
   and Virginia Tech - Wake Forest University School of Biomedical Engineering and Sciences,
   Virginia Polytechnic Institute and State University,
   Blacksburg, VA 24061, USA.

4 Department of Biochemistry,
   Boston University School of Medicine,
   Boston, MA 02118, USA.

**Running Title:** Clustering Effect on Ligand Dissociation

[4] Corresponding author: e-mail: kfw@vt.edu
[5] Present address

## Abstract

Receptor-ligand binding is a critical first step in signal transduction and the duration of the interaction can impact signal generation. In mammalian cells, clustering of receptors may be facilitated by heterogeneous zones of lipids, known as lipid rafts. *In vitro* experiments show that disruption of rafts significantly alters the dissociation of fibroblast growth factor-2 (FGF-2) from heparan sulfate proteoglycans (HSPG), co-receptors for FGF-2. In this paper, we develop a continuum stochastic formalism in order to address how receptor clustering might influence ligand rebinding. We find that clusters reduce the effective dissociation rate dramatically when the clusters are dense and the overall surface density of receptors is low. The effect is much less pronounced in the case of high receptor density and shows non-monotonic behavior with time. These predictions are verified via lattice Monte Carlo simulations. Comparison with FGF-2-HSPG experimental results is made and suggests that the theory could be used to analyze similar biological systems. We further present an analysis of an additional co-operative "internal diffusion" model that might be used by other systems to increase ligand retention when simple rebinding is insufficient.

# (I) INTRODUCTION

The cell membrane is composed of many different types of lipid species. This heterogeneity leads to the possibility of organization of different species into distinct *domains* (1). Such domains are especially suited and designed for specialized functions such as signal transduction, nutrient adsorption, and endocytosis. They can link specific cellular machinery and physical features and are equipped with mechanisms for maintenance (addition and removal of specific molecules) for a certain period of time, during which the domains may diffuse as single entities (2). Lipid rafts, which are micro-domains rich in sphingolipids and cholesterol, represent one of the most interesting but insufficiently understood lipid domains (3). Various estimates are available for raft sizes, and diameters in the range 25-200 nm have been reported using various methods (4). A limitation in this area remains that the definition of lipid rafts is rather broad and currently includes a wide range of what will likely prove to be distinct domains that may be distinguished by the particular protein and lipid compositions (2,4,5). Operational definitions of rafts based on resistance to detergent solubilization and sensitivity to cholesterol removal are limited by artifacts of the various procedures used to define rafts and on difficulties in relating model membranes to cell membranes. Nonetheless, it is clear that cell membranes are not homogeneous and that protein-protein, protein-lipid and lipid-lipid interactions all participate in regulating raft size, dynamics and function. Consequently, a myriad of functions have been prescribed to lipid rafts; one possibility being that lipid rafts may serve as mediators of signal transduction for several growth factors, including fibroblast growth factor-2 (FGF-2) (6-8).

Growth factors act as triggers for many cellular processes and their actions are typically mediated by binding of ligand to the extracellular domain of transmembrane receptor proteins. For many receptors, signal transduction requires dimerization or clustering whereby two or more receptors, following ligand binding, interact directly to facilitate signal transduction. While ligand binding is generally specific to members of a family of transmembrane receptor proteins, heparin-binding growth factors such as FGF-2 interact with both specific members of the FGF receptor family and heparan sulfate glycosaminoglycan chains of cell surface proteoglycans (HSPG). HSPG represent a varied class of molecules, including the transmembrane syndecans, the glycosyl-phosphotidylinositol anchored glypicans, and extracellular proteoglycans such as perlecan (reviewed in 9,10). The interaction of FGF-2 with HSPG is of a lower affinity than to the cell surface signaling receptor (CSR) but has been shown to stabilize FGF-2-CSR binding and activation of CSR (11,12). Moreover, HSPG have recently been demonstrated to function directly as signaling receptors in response to FGF-2 binding, leading to the activation of protein kinase C alpha (12) and Erk1/2 (6).

There is evidence that cell surface HSPG are not distributed uniformly, but are instead localized in lipid rafts (6,14-16), and this association may be facilitated by FGF-2 binding and clustering (17). This localization and clustering may further have a dramatic influence on signaling through both persistence of signaling complexes and localization with intracellular signaling partners. For example, FGF-2 dissociation kinetics from HSPG were significantly altered when cells were treated with the lipid raft-disrupting agents methyl-β-cyclodextrin (MβCD) (Fig 1). Retention of FGF-2, even at long times, was significantly greater in the untreated state, suggesting that rafts regulate this process. These experiments suggest that clustering of HSPG in lipid rafts effectively slows down dissociation by increasing the rebinding of released FGF-2. If this is indeed true, then the localization of binding sites to micro-domains on the cell surface could be an important mechanism employed by receptors to boost signal transduction via increased persistence.

The relation between the apparent association and dissociation rates of ligands interacting with receptors on a (spherical) cell surface with the corresponding intrinsic rates has been studied previously by several authors (18-23). Berg and Purcell (18) demonstrated that for ligands irreversibly binding to N receptors on a spherical cell of radius $a$, the effective forward rate constant becomes a non-linear function of N, assuming the form $k_f = 4\pi D a [N k_+ / (4\pi D a + N k_+)]$, where $k_+$ is the association rate for a single receptor in close proximity to the ligands (i.e., the intrinsic binding rate). The quantity in brackets was termed the 'capture probability' $\gamma$ by Shoup and Szabo (19). The effective dissociation rate was analogously defined as the product of the intrinsic rate and the 'escape probability' $1-\gamma$. This leads to (19, 24)

$$k_r = k_- \left[ \frac{k_D}{k_D + N k_+} \right] , \qquad (1)$$

where $k_-$ is the intrinsic dissociation rate and $k_D$ ($=4\pi D a$ for a spherical completely absorbing surface) represents the diffusion-limited association rate, illustrating how increasing receptor numbers lead, in general, to a decrease in apparent dissociation rate. This result was later extended to include the presence of solution receptors by Goldstein et al. (24). Association of ligands to a cluster of receptors on a planar surface, which also includes the surface diffusion of ligands inside the cluster, was investigated by Potanin et al. (26). This study predicted a non-monotonic variation of the forward rate constant with cluster size that was found to fit better with some experimental results.

In general, the effective dissociation rate of ligands from a set of receptors depends on the frequency of rebinding, whereby a dissociated ligand wanders around in the solution for some time and reattaches to the binding surface upon contact. This is only implicitly included in the above approaches. A systematic mathematical study of the rebinding probability of a single ligand was undertaken by Lagerholm and Thompson (27). An independent self-consistent mean-field model of rebinding of ligands bound to receptors in an infinite two-dimensional plane was recently presented by us (28) in the context of analyzing Surface Plasmon Resonance (SPR) experiments.

In the present paper, we generalize our earlier discrete model (28) to incorporate a continuum description for the receptor distribution as well as the ligand motion. The self-consistent stochastic mean-field theory of rebinding thus developed is then used as the basis for extending our investigation to include non-uniformity in the spatial distribution of receptors. In particular, we study how rebinding is affected by the presence of receptor clusters on the cell surface. Our broad conclusions from this study are as follows: (i) Receptor clustering dramatically reduces the effective dissociation rate through enhancing rebinding, if the overall receptor density is small enough that the effect would have been negligible without clustering. (ii) When the overall receptor density is high, the effect of clustering is smaller, but the frequent rebinding events render the dissociation non-exponential in the case of a planar surface.

In the remainder of this paper, we first develop the theoretical formalism to study rebinding of ligands to an infinite plane of uniformly distributed receptors. Motivated by recent experimental observations of the effect of lipid rafts on ligand rebinding (6), the formalism is then extended to include receptor clusters. Subsequently, our theoretical predictions are compared to Monte Carlo simulation data. Finally, we comment on possible applications, including a possible "internal diffusion" model extension, and discuss consequences for the analysis of experimental results.

## (II) THEORY

### (i) REBINDING ON A PLANAR SURFACE

In this section, we present a generalization of our recently introduced lattice random walk based theory of rebinding to a continuum distribution of receptors on a two-dimensional infinite surface. Let us consider a homogeneous distribution of receptors on an infinite planar surface with constant mean surface density $R_0$ per unit area. The intrinsic dissociation and association rates are denoted by $k_-$ and $k_+$, respectively. We denote by $R(t)$ the density of receptors bound to the ligand at any time t, and its dynamical equation has the form

$$\frac{dR(t)}{dt} = -k_- R(t) + k_+ \rho(t)[R_0 - R(t)] \quad , \tag{2}$$

where $\rho(t)$ represents the ligand density in the vicinity of the surface. Let us now consider a dissociation experiment for which the density of receptors that are bound to ligands at time $t=0$ is $R(0) = R^*$, and the ligand density in the bulk volume is taken to be zero at $t=0$. It then follows that a non-zero $\rho(t)$ at time $t>0$ is entirely due to ligands released from bound receptors at previous times $0 < \tau < t$. Taking this previous history carefully into account, we may write down an expression for $\rho(t)$ in the concise following form:

$$\rho(t) = R_0 k_- \int_0^t d\tau\, p(\tau) C_{R_0}(t-\tau) \int d^2 r\, G_2(r, t-\tau) \quad , \tag{3}$$

where
(i) $C_{R_0}(t)$ denotes the (surface-integrated) one-dimensional probability density (with dimension of 1/length) of a random walk returning to its point of origin at time $t$, given that the origin constitutes a partially absorbing barrier with a density $R_0 - R(t)$ of absorbing points per unit area, and
(ii) $G_2(r,t) = [4\pi Dt]^{-1} \exp(-r^2/4Dt)$ represents the (normalized) two-dimensional probability density for finding a diffusing particle at distance $r$ from the origin at time $t$.

In order to eliminate the time dependence of the boundary condition in (i), we choose $R^* \ll R_0$. Let $p(t) = R(t)/R_0$ be the fraction of receptors bound to ligands at time $t$, so that $p(0) = R^*/R_0 \ll 1$ (which also implies $p(t) \ll 1$). When the spatial integration in Eq.3 is extended to infinity, Eq.2 is thus reduced to

$$\frac{dp(t)}{dt} = -k_-\left[p(t) - k_+ R_0 \int_0^t d\tau\, p(\tau) C_{R_0}(t-\tau)\right] \quad . \tag{4}$$

We note that the rebinding problem as defined by Eq.4 is effectively one-dimensional i.e., the two in-plane dimensions have disappeared upon spatial integration. This feature enables many important simplifications, as will be obvious from the following discussions.

The quantity $C_{R_0}(t)$ is now calculated from the frequency of first passage events: Let $q(\tau)$ denote the probability density of ligands that at time $\tau$ return to the surface for the first time following dissociation. At this point in time, the ligands may be either absorbed or reflected back to the solution and subsequently return at a later time. $C_{R_0}(t)$ could then be calculated by summing over of all such events.

In order to proceed with our formalism, it is useful to imagine the available space to be divided into cubic elements (i.e., 'coarse-grain' the space), each with volume $\lambda^3$. Here $\lambda$ is a 'coarse-grained' length scale, which we assume to be of the order of the size of a single ligand molecule. The ligand diffusion may now be viewed effectively as transfer of its center of mass between such elements. When a ligand occupies an element of volume adjacent to the surface, it may become bound to a receptor, and the probability for this to occur is denoted $1-\gamma$, so that $\gamma$ is the probability of non-absorption of the ligand upon encounter. The equation for $C_{R_0}(t)$ thus satisfies the integral equation

$$C_{R_0}(t) = q(t) + \gamma \frac{\lambda}{2\delta} \int_{2\delta}^t d\tau\, q(\tau) C_{R_0}(t-\tau) \quad . \tag{5}$$

The factor $2\delta$ is the smallest time over which a rebinding event can take place: $\delta$ is a microscopic time scale, which is the interval between two successive collisions of the ligand and the solvent molecule (which, for simplicity, we assume to be a non-fluctuating constant), which cause the ligand to move away from the surface [6]. In order to solve the integral equation (5), we express it in terms of the Laplace-transformed variables $\tilde{f}(s) = \int_0^\infty dt\, e^{-st} f(t)$, whence we obtain (in the limits $\delta \to 0$, $\lambda \to 0$, with $\lambda/2\delta$ held fixed):

---

[6] In general, $\delta$ is independent of the coarse-graining length scale $\lambda$. However, if we approximate the ligand diffusion as a discrete 'random walk' (as in simulations), then these are related as $\delta = \lambda^2/2D$.

$$\tilde{C}(s) = \frac{\tilde{q}(s)}{1 - \gamma \frac{\lambda}{2\delta} \tilde{q}(s)} \quad . \tag{6}$$

The calculation of $\tilde{q}(s)$ is outlined in Appendix A, with the result

$$\tilde{q}(s) = \frac{1}{\sqrt{Ds} + \frac{\lambda}{2\delta}} \quad . \tag{7}$$

(Throughout this paper, we will define $D$ to be the diffusion coefficient of the random walk projected onto the z-axis, perpendicular to the plane containing receptors. Its relation to the complete three-dimensional diffusion coefficient $D^*$ is simply $D = (1/3)D^*$).

The Laplace-transformed version of Eq.4 after all the above substitutions reads

$$\tilde{p}(s) = \frac{p(0)}{s + k_-[1 - \Sigma(s)]} \quad , \quad \text{with} \quad \Sigma(s) = k_+ R_0 \tilde{C}(s) \quad . \tag{8}$$

The probability of absorption upon encounter (which we denoted $1-\gamma$) may be computed as follows: Consider a ligand molecule very close to the surface, occupying a cell of volume $\lambda^3$. The ligand density in its vicinity is $\rho = \lambda^{-3}$. The probability that there is a receptor within the adjacent surface area $\lambda^2$ is $R_0 \lambda^2$. The ligand stays close to the surface for a time interval $2\delta$ (since no diffusion is possible through the surface), so that the probability of binding during this time interval is $1 - \gamma \equiv \alpha = k_+ \rho \times R_0 \lambda^2 \times 2\delta = 2k_+ R_0 \delta / \lambda$.[7]

After substitution of Eqs.6 and 7 into 8, and employing the above result to substitute for $\gamma$, we arrive at

$$\Sigma(s) = k_+ R_0 / \left[\sqrt{Ds} + k_+ R_0\right] \quad , \tag{9}$$

and consequently

$$\tilde{p}(s) = \frac{p(0)}{s + k_- \left(\frac{\sqrt{Ds}}{\sqrt{Ds} + k_+ R_0}\right)} \quad . \tag{10}$$

Let us now seek to extract the time dependence of the fraction $p(t)$ from this expression. Clearly, at short times, i.e., when $s$ is sufficiently large, namely $s \gg (k_+ R_0)^2 / D$, $\tilde{p}(s) \approx p(0)/(s + k_-)$, and the decay is purely exponential with the intrinsic rate constant $k_-$. In this early-time regime, rebinding does not yet occur. On the other hand, in the very late time regime corresponding to $s \ll (k_+ R_0)^2 / D$, we have $\tilde{p}(s) \approx p(0)/[s + k_- \sqrt{Ds}/(k_+ R_0)]$. The explicit time dependence of $p(t)$, therefore, exhibits the limiting behaviors after Laplace inversion:

$$p(t) \approx p(0) exp(-k_- t) \quad \text{for} \quad t << \frac{D}{(k_+ R_0)^2} \equiv t_e \quad , \tag{11a}$$

$$p(t) \approx p(0) e^{ct} erfc(\sqrt{ct}) \quad \text{for} \quad t >> \frac{D}{(k_+ R_0)^2} \quad , \tag{11b}$$

where $c = D\left(\frac{K_D}{R_0}\right)^2$ and $K_D = k_-/k_+$ is the equilibrium dissociation constant. Our self-consistent mean-field analysis thus yields that the ligand dissociation curve for a planar surface is always non-exponential for sufficiently late times. However, for small association rate or surface coverage, the initial transient regime showing exponential decay could well last for substantial durations.

---

[7] Since the absorption probability $\alpha \leq 1$, this implies that the product $k_+ \delta$ must be bounded from above. However, since $\delta$ is sufficiently small ($10^{-10} - 10^{-12}$ sec), this is hardly ever an issue, even for very high values of $k_+$.

The non-exponential decay in Eq.11 was also predicted in a previous lattice model of the problem developed to model SPR experiments (28). Indeed, one can show that with the appropriate mapping, the time constants c of the continuum and the lattice models coincide. In the discrete variant, the receptors are distributed on a lattice (with unit length ) at a mean density $\theta_s$, and upon hitting a receptor (the sizes of both ligand and receptor are assumed negligible in comparison with ), a ligand is absorbed with probability $\theta_a$. The 'effective' surface coverage is therefore given by $\theta = \theta_s \theta_a$. These parameters are related to the continuum variables through the relations $R_0 = \theta_s / {}^2$ and $k_+ = \theta_a D$ . Upon making these substitutions in Eq.11, we find that the expressions corresponding to the continuum and lattice formalisms match perfectly.

## (ii) EXTENSION TO RECEPTOR CLUSTERS

In this section, we adapt the stochastic self-consistent mean-field theory for ligand rebinding presented above to incorporate non-uniform spatial receptor distributions. We consider receptors distributed in clusters of radius $r_0$, such that the density of receptors inside the clusters is $R'_0 > R_0$, where the latter represents the mean density of receptors on the surface.

In order to generalize the previous theory to incorporate receptor clusters, we adopt the following approximation: Any rebinding event where the originating and the final points are separated by a distance $r < \xi$ is assumed to take place in a local environment with receptor density $R'_0$, whereas any ligand that travels a lateral distance $r \geq \xi$ to rebind is assumed to sense only a smaller receptor density $R_0^*$ that we assume to be of the order of the mean density $R_0$. In order for this approximation to be useful, we need to identify $\xi$ with a physical length scale: here we simply assume that $\xi \sim r_0$. It must be noted that no strict spatial segregation exists between the two classes of rebinding events in the actual system. However, it will be seen later in comparison with numerical results that this approximation is remarkably successful in predicting the different temporal decay regimes in the presence of receptor clusters.

We shall now quantify these ideas using the previously developed formalism as a basis. The complete expression describing the dynamics of the bound fraction, which obviously generalizes Eq.4, becomes

$$\frac{dp(t)}{dt} = -k_-\left[p(t) - k_+\int_0^t d\tau \int_0^\infty d^2r R(r) G_2(r, t-\tau) C_{R(r)}(t-\tau) p(\tau)\right] \quad , \tag{12}$$

where, according to our earlier assumptions, the distance-dependent coverage function $R(r)$ assumes the step function form $R(r) = R_0' + (R_0 - R_0')\Theta(r - \xi)$, where $\Theta(x)$ denotes the Heaviside step function, with the properties $\Theta(x) = 0$ when $x < 0$ and $\Theta(x) = 1$ when $x \geq 0$.

Let us consider two special cases of interest:

**Case 1: Dense isolated clusters on a planar or spherical surface: $R_0 \approx 0$; $R_0'$ large.**
This situation is realized when the clusters are tightly packed with receptors, but the number of clusters themselves is small, so that the mean surface coverage of receptors is a negligible fraction. In this case, the homogeneous part of the rebinding term in Eq.12 is vanishingly small, and the equation reduces to

$$\frac{dp(t)}{dt} = -k_-\left[p(t) - k_+ R_0'\int_0^t d\tau p(\tau) C_{R_0'}(t-\tau)\left(1 - e^{-\frac{\xi^2}{4D(t-\tau)}}\right) - O(R_0)\right] \quad . \tag{13}$$

It is important to note that Eq.13 is valid also for receptor clusters on a spherical cell surface, provided the size of the cluster $\xi$ is much smaller than the radius of the cell itself. The Laplace transform of $p(t)$ has the form of Eq.8, with

$$\Sigma(s) \approx k_+ R_0' \int_0^\infty dt e^{-st} C'(t)\left[1 - e^{-\frac{\xi^2}{4Dt}}\right] \quad , \tag{14}$$

where we have introduced the concise notation $C'(t) = C_{R_0'}(t)$ (and similarly $C(t) \equiv C_{R_0}(t)$ in future calculations). In order to evaluate the Laplace transform of the function $C'(t)\exp(-\xi^2/4Dt)$, we apply the following trick: Using Eq.B2 in Appendix B for the limiting forms of the function $C'(t)$, we approximate it as

$$C'(t) \approx \Theta(t_0 - t)(\pi Dt)^{-1/2} + \Theta(t - t_0)\sqrt{\frac{D}{4\pi}}(k_+ R_0')^{-2} t^{-3/2} \quad , \text{ where } t_0 \approx D(k_+ R_0')^{-2} \quad , \tag{15}$$

and $\Theta(x)$ again is the Heaviside step function.

We now substitute this expression into Eq.14, and use it to evaluate the $\xi$-dependent term in the brackets. (The first term gives $\left[\sqrt{Ds} + k_+ R'_0\right]^{-1}$, see Eq.9.) After inserting the result $\int_0^\infty dt\, t^{-3/2} e^{-\xi^2/4Dt - st} = \xi^{-1}\sqrt{4\pi D}\, e^{-\xi\sqrt{s/D}}$ (29), we arrive at the following expression:

$$\frac{\Sigma(s)}{k_+ R'_0} \approx \frac{1}{\sqrt{Ds} + k_+ R'_0} - \int_0^{t_0} dt \left[\frac{1}{\sqrt{\pi Dt}} - \sqrt{\frac{D}{4\pi}}(k_+ R'_0)^{-2} t^{-3/2}\right] e^{-st - \xi^2/4Dt} - \frac{D}{(k_+ R'_0)^2 \xi} e^{-\sqrt{\xi^2 s/D}}.$$
(16)

In particular, we are interested in the long-time limit $t \gg \xi^2/4D$ (i.e., times much larger than the typical time scale for ligand diffusion across a cluster), corresponding to $s \ll 4D/\xi^2$. In this limit, the rebinding term has the form $\Sigma(s) \approx \Sigma(0) + O\!\left(\sqrt{s\xi^2/D}\right)$, with:

$$\Sigma(0) \approx 1 - \frac{1}{\sqrt{\pi}} \frac{\xi}{\xi_0} \Gamma\!\left[-\frac{1}{2}, \left(\frac{\xi}{\xi_0}\right)^2\right] - \frac{\xi_0}{2\xi}\left\{1 - \frac{1}{\sqrt{\pi}} \Gamma\!\left[\frac{1}{2}, \left(\frac{\xi}{\xi_0}\right)^2\right]\right\},$$
(17a)

where we have defined the length scale

$$\xi_0 = \frac{2D}{k_+ R'_0},$$
(17b)

and $\Gamma(a,x) = \int_x^\infty dy\, y^{a-1} e^{-y}$ represents the incomplete Gamma function (30).

Let us now assume that the clusters are very densely packed with receptors, i.e., $R'_0$ is sufficiently large so that $\xi \gg \xi_0$. In this case, the contributions in Eq.17a that involve incomplete Gamma functions are small ($\Gamma(a,x) \sim x^{a-1} e^{-x}$ for $x \gg 1$, (30)). Therefore, $\Sigma(0) \approx 1 - \xi_0/2\xi$ in Eq.17a when $\xi \gg \xi_0$. After substitution in Eq.8, we see that

$$\tilde{p}(s) \approx \frac{p(0)}{s + k_- \xi_0/2\xi}$$
(18)

as $s \to 0$. After Laplace inversion,

$$p(t) \approx p(0) \exp\!\left(-k_- \frac{\xi_0}{2\xi} t\right) \quad \text{for } t \gg \xi^2/D \text{ and } \xi \gg \xi_0.$$
(19)

From Eq.19, the length scale $\xi_0$ (or, more precisely, $\xi_0/2$) may be understood as the threshold size a cluster needs to have in order to appreciably affect the dissociation.

We thus reach an intriguing conclusion: When the mean surface density is sufficiently small, clustering of receptors has (over sufficiently long time scales) the effect of *reducing* the effective dissociation rate of ligands by a factor that is inversely proportional to the size of the cluster. It should also be borne in mind that the very late time regime for *any* small but non-zero mean density should display the non-exponential behavior of Eq.11b. However, the characteristic time scale for entry into this regime (for a uniform distribution) grows as $R_0^{-2}$, and is likely to be masked by other effects (e.g., finite-size effects, non-specific binding) in experiments.

In order to view this result in the context of the previous findings of Berg and Purcell (18) and Shoup and Szabo (19), we may compare Eq.19 with the analogous result in Eq.1 obtained via very different arguments. Let us imagine that the density of receptors inside a cluster is so high (consistent with our own assumptions in reaching Eq.19) that the cluster effectively acts like an absorbing disk, for which the diffusion-limited onward rate constant is $k_D = 4Dr$ (31) where $r$ is the radius of the cluster. Let $N$ be the number of receptors inside a cluster, which we assume to be so large that $Nk_+ >> k_D$ in the denominator of Eq.1. After re-expressing $N$ in terms of the receptor surface density $R_0' = N/\pi r^2$, we find that, within this approximation, the reduction factor for the association rate in Eq.1 is identical to that in Eq.19, with $\xi = (\pi/4)r$, an aesthetically pleasing result. It should, however, be emphasized that the framework of our theory is more general and provides a broader perspective.

When the radius $\xi$ is sufficiently large ($\xi >> \xi_0$), there is also another (intermediate) time regime $D(k_+ R_0')^{-2} << t << \xi^2/D$, for which the last term in Eq.16 is small, and the first term dominates (since again the incomplete Gamma functions vanish in the limit of large $\xi$ specified above). In this regime, we hence recover the non-exponential dissociation encountered earlier in the section 'Rebinding on a planar surface' in the context of a homogeneous receptor distribution, see Eq.(11b):

$$p(t) \approx p(0)e^{\tilde{c}t} erfc(\sqrt{\tilde{c}t}) \text{ , when } D(k_+ R_0')^{-2} << t << \xi^2/D; \quad \tilde{c} = D(K_D/R_0')^2. \quad (20)$$

In this intermediate time regime, the ligand behaves as if diffusing in the presence of an infinite substrate with receptor density $R_0'$.

The preceding calculations, in particular Eq.17-19, show that the clusters have to be of a minimum size ($\sim \xi_0 = 2D(k_+ R_0')^{-1}$) if they are to produce a significant effect on the dissociation. It is, therefore, important to know how this cut-off size compares with independent estimates for the size of lipid rafts. The total number of proteins likely to be contained inside a raft of area $2100 nm^2$ has been estimated to be in the range 55-65 (32), assuming very close packing, or close to 20 (33) assuming the same density of packing inside the raft and the surrounding membrane. The number of specific proteins like HSPG is possibly less. As a conservative estimate, we assume that there are $n \sim 5-10$ HSPG inside a raft, which gives $R_0' = n/\pi r^2$, where r is the raft radius. The condition that clusters affect dissociation substantially is $\xi/\xi_0 \geq 1$, from our previous analysis. Let us now make the identification $\xi \approx r$, which, combined with the previous estimate for the receptor density, gives the condition $k_+ \geq 2\pi Dr/n$. Let us use $r \sim 25$ nm as a rough estimate for the size of a lipid raft (34), which then gives $k_+ \geq 10^8 - 10^9 M^{-1}s^{-1}$, if we assume a diffusion coefficient $D = 10^{-10} m^2 s^{-1}$.

Our conclusion, therefore, is that rafts of extensions in the range 25-50 nm should be capable of producing a measurable effect on ligand dissociation purely by a diffusion-controlled mechanism, provided the association rate of the specific protein is large enough. It must, however, be remarked that this conclusion strictly applies to monovalent ligands interacting with a monovalent single receptor only. If, as in the specific case of FGF-2, there is more than one receptor that can bind the ligand and the possibility of higher order complexes exists, then the inclusion of surface biochemical coupling reactions needs to be taken into account. In the section 'Comparison with experiments', we provide a more detailed discussion of these aspects in the context of experiments with HSPG.

**Case 2: High mean surface density: perturbation theory for small rafts**
When the mean surface density of receptors is high, one might expect that rebinding has significant effect on dissociation even without any additional clustering mechanisms and that any effect of rafts on dissociation would be confined to sufficiently small time scales. This argument is, in fact, supported by numerical simulations that we present below. Yet here we aim to quantify the effect of clustering on ligand rebinding in the case of high mean surface density. For this purpose, Eq.12 is conveniently rewritten in the form

$$\frac{dp(t)}{dt} = -k_-\left[p(t) - k_+ R_0 \int_0^t d\tau p(t-\tau) C(\tau) - k_+ \int_0^t d\tau p(t-\tau)\left(1 - e^{-\frac{\xi^2}{4D\tau}}\right)\left[R_0' C'(\tau) - R_0 C(\tau)\right]\right], \quad (21)$$

where $R_0' \geq R_0$. The second term inside the brackets is the homogeneous rebinding term, whereas the third term is the correction term arising from clustering. We observe that, for any fixed $\xi$, the latter term gets progressively smaller at sufficiently large times, which implies that the late time behavior must be dominated by the homogeneous term. In order to make further analytic progress, let us now focus on the regime of small clusters, with $\xi \ll \xi_0 = 2D(k_+ R_0')^{-1}$. We may then use the small-time (surface-density independent) form for the functions $C$ and $C'$ from Eq. B2a in Eq.21. It follows that the effective equation for $p(t)$ (over short times) has the form

$$\frac{dp(t)}{dt} = -k_- p(t) + k_- k_+\left[R_0 \int_0^t d\tau C(\tau) p(t-\tau) + (R_0' - R_0)\int_0^t d\tau \frac{1 - e^{-\frac{\xi^2}{4D\tau}}}{\sqrt{\pi D \tau}} p(t-\tau)\right], \quad (22)$$

where the last term is the correction due to the presence of clusters. Note that the correction term vanishes when $R_0' = R_0$ and $\xi = 0$. Eq.22 is valid only for sufficiently small times $t \ll \xi^2/D$, as the replacement of the functions $C$ and $C'$ by the surface density-independent form (Eq.B2a) gets progressively more inaccurate at larger and larger times. This equation is also solved using the Laplace transform technique, and the cluster correction term (as defined in Eq.8) is found to have the form

$$\Sigma(s) = k_+ R_0 \tilde{C}(s) + \frac{k_+(R_0' - R_0)}{\sqrt{Ds}}\left[1 - e^{-\xi\sqrt{\frac{s}{D}}}\right], \quad \xi \ll \xi_0 \quad . \quad (23)$$

After substituting in Eq.8, we obtain, for $t >> D(k_+ R_0')^{-2}$:

$$\tilde{p}(s) = \frac{p(0)}{s + \sqrt{cs} - k_- \omega} \quad , \quad \text{where } \omega = 2\varepsilon(\xi/\xi_0) \tag{24}$$

and $\varepsilon = 1 - R_0 / R_0'$. Eq. 24 holds in the time interval where the last term in Eq.24 is small compared with the first, and the regime of validity thus turns out to be $t << t' \approx (k_- \omega)^{-1}$. In accordance with our earlier assumption on the cluster size, $\omega$ is now a small (dimensionless) parameter, and this enables the expression in Eq.(24) to be expanded as a perturbation series (which would require that $s$ is sufficiently large, or equivalently, we need to restrict ourselves to sufficiently small times) of the form

$$\tilde{p}(s) = \frac{p(0)}{s + \sqrt{cs}} + \frac{p(0)}{[s + \sqrt{cs}]^2} k_- \omega + O(\omega^2) \quad . \tag{25}$$

We may now write $p(t) = p_0(t) + \hat{p}(t)$, where $p_0(t)$ is given by Eq.11b and $\hat{p}(t)$ is the cluster-correction term that is determined by inverting the second term in Eq.25. The complete expression is

$$\hat{p}(t) = k_- \omega p(0) \left[ c^{-1} + (2t - c^{-1}) e^{ct} \operatorname{erfc}(\sqrt{ct}) - 2\sqrt{t/\pi c} \right] + O(\omega^2) \quad , \tag{26}$$

where the constant c was defined following Eq.11. Eq.26 provides the first correction term in the bound fraction, for small clusters. As will be seen in the next section, this expression reproduces the cluster correction term in simulations approximately, but only at early times (which is consistent with our own assumption that t should be sufficiently small).

*To summarize this section*, the theoretical formalism we have presented predicts a number of interesting regimes for the effective dissociation of ligand from receptors on cell surfaces. For a uniformly distributed set of receptors on a plane, we find that the decay is exponential with the intrinsic dissociation rate initially (Eq.11a), but crosses over to a non-exponential decay at later times (Eq.11b) owing to multiple rebinding events. When the receptors are clustered, the effects of rebinding depend on the mean receptor density. When the mean density is low so that no appreciable rebinding occurs with a uniform distribution, clustering is predicted to have the effect of producing an exponential decay at intermediate times with a reduced decay coefficient that is a function of the cluster size and the other parameters (Eq.19). The very late time behavior is still presumably non-exponential, although a full characterization of this crossover has not yet been performed. When the mean density is sufficiently high, the effect of clustering was found to be non-monotonic, small at early and late times and reaching a maximum at a certain intermediate time.

In order to check our analytical results, in particular Eqs.19 and 26, we have performed lattice Monte Carlo simulations, which will be the subject of the next section.

## (III) RESULTS

### (i) LATTICE MONTE CARLO SIMULATIONS

The 'hopping between elements' picture of ligand diffusion we presented in the section 'Rebinding on a planar surface' is easily implemented in numerical simulations. The substrate surface is envisioned as a two-dimensional square lattice, with the length scale $\lambda$ setting the lattice spacing. The unit time scale is set to $\delta t = \lambda^2/2D$, the time scale of hopping between elements. (We use a different symbol here to distinguish from the more fundamental time scale $\delta$ introduced in the section 'Rebinding on a planar surface'.) Using these units, all quantities we discussed above may be expressed in dimensionless form (see Table 2). The ligand motion is modeled as a three-dimensional random walk between elements in the space above the substrate.

In the simulations, we choose the association rate to be $k_+ = D\lambda$. With this choice, the binding rate of the ligand close to a receptor is $p = \lambda^{-3}k_+ = D\lambda^{-2}$ and the probability of binding over a single Monte Carlo time step for a ligand close to the surface is $\tilde{k}_+ = p\delta t = 1/2$, i.e., the binding is purely diffusion-limited. In real units, this choice corresponds to an association rate of $\sim 10^{-13} cm^3 s^{-1} \sim 10^6 M^{-1} s^{-1}$. A smaller value of $k_+$ involves only a trivial modification of the algorithm: The probability of binding is reduced to $\tilde{k}_+ = k_+/D\lambda$ (in simulations, this factor may be simply absorbed into the dimensionless surface coverage, while keeping the binding purely diffusion-limited), but a larger association rate would require a more microscopic simulation, and is not addressed in this paper.

We next discuss our choice for the dissociation rate. A realistic value of $k_-$ would fall in the range of $10 - 10^{-4} s^{-1}$, which means that the dimensionless rate $\tilde{k}_- = k_- \delta t$ would be a very small number (For $\lambda \approx 5nm$ and $D \sim 10^{-10} m^2 s^{-1}$, we estimate $\delta t \sim 10^{-7} s$), of the order of $10^{-6} - 10^{-11}$. Since the time scale of measurement of dissociation would have to be at least of the order of $k_-^{-1}$, this would require the simulation to be run over $\tilde{k}_-^{-1}$ Monte Carlo steps. For computational efficiency, therefore, we choose $\tilde{k}_- = 10^{-4}$ in all the simulations.

The surface density of receptors $R_0$ is the next important parameter in the model, and its dimensionless version is denoted by $\theta = R_0 \lambda^2$. Assuming that the ligands and the receptor extracellular binding domains are not significantly different in size, the range of allowed values for this parameter is $\theta \leq 1$. In the substrate lattice, therefore, $\theta$ simply represents the fraction of binding sites. Note that the simulations also could correspond to the case where the association rate $k_+ < D\lambda$, where we would maintain the binding to be diffusion-limited, but effectively reduce $\theta$ to $\theta' = \theta(k_+/D\lambda)$ in the simulation runs.

Our strategy is as follows: Keeping the overall density $\theta$ constant, we arrange the receptors into N clusters of (dimensionless) radius $\tilde{r}_0 \geq 1$. Because of lattice constraints, it is not possible to ensure that all the receptors are contained in such clusters. Rather, our criterion is that, for a certain value of $\tilde{r}_0$, N be selected such that the number of receptors outside clusters is kept a minimum. The simulations are done with reasonably large lattices ($10^3 \times 10^3$) so that small surface coverage could be explored. Fig.2 shows two typical receptor configurations used in our simulations. All the data was averaged over 100 different initial realizations of the receptor configuration.

The ligand diffusion is governed by periodic boundary conditions on the four borders of the lattice so that a ligand that exits at one boundary reenters from the opposite side. The direction perpendicular to the plane of the lattice shall be referred to as the z-axis, and the surface itself is located at z=0. The ligand diffusion in the z-direction is not upper bounded. We also neglect surface diffusion of the receptor proteins, irrespective of their being clustered or isolated, and treat them as static objects throughout this paper (see, however, the discussion at the end of this section). At the beginning of the dynamics, a fraction $p(0)$ of all the receptor sites are bound to a single ligand each. Although the precise value of p(0) is unlikely to have a large impact on the late-time decay, we chose p(0)=0.25 in all the simulations so that we are not too far from the approximation p(0) << 1 made in the set-up of the theory.

There are three main dynamical processes in the simulation: (i) Dissociation of a ligand from a bound receptor takes place with probability $\tilde{k}_- = k_- \delta t$ per time step $\delta t$. This move updates the position of the ligand from z=0 to z=2, in units of the lattice spacing. (We use z=2 instead of z=1 in order to prevent immediate rebinding to the same receptor.) (ii) Diffusion of the released ligands in solution: A free ligand moves a distance equal to one lattice spacing in one of the six directions with probability 1/6 per time step. (iii) Re-adsorption of free ligands to free receptors: A free ligand at z=1 is absorbed by a free receptor below it, if there is one, with probability 1.

Our initial simulations were done at two values of the surface coverage ($\theta = 10^{-3}$ and $\theta = 10^{-1}$) and we find that the surface density has a dramatic impact on the dissociation rate (Fig.3). The first case ($\theta = 10^{-3}$) corresponds to very sparsely distributed receptors, whereas the distribution is quite dense in the second case ($\theta = 10^{-1}$). As shown, the decay at the low density appears exponential and has a measured decay constant of $\sim 0.7 \times 10^{-4}$, approximately reflective of the true dissociation rate ($\tilde{k}_- = 10^{-4}$). For the more dense system, a distinctly non-exponential decay is evident. However, a closer look shows that at early times (t < 200 Monte Carlo steps) an exponential decay for the high coverage case also, in accordance with Eq.11a (Fig.3 - inset), can be found. The decay constant measured in the simulations by fitting this early part (t 400 Monte Carlo steps from Eq. 11a) to an exponential curve is close but somewhat lower than the intrinsic rate used for the simulations ($\sim 0.6 k_-$) which we believe is simply an artifact of the discrete algorithm used in the simulations: In Appendix C, we show that the effective decay constant in the case of even a single isolated receptor and a ligand in a three-dimensional cubic lattice (such as used in our simulations) is less than the intrinsic rate, on account of the non-zero return probability of the lattice random walk. The non-exponential curve for the high-density case fits well with the theoretical prediction in Eq.11b (which has also been supported by dissociation data from surface plasmon resonance experiments in a recent study (27). Note that in both the low and high mean density cases the simulations were set up so that the clusters were completely full of receptors (i.e., with the highest density possible in those regions). Also, as noted above, the low mean surface density could also correspond to the case where the association rate is low ($k_+ < D\lambda$).

We next addressed how clusters might impact dissociation focusing first on the low-coverage regime. The coverage we chose was $\theta = 10^{-3}$ (in terms of distribution over the cellular surface, this would roughly correspond to ~ $10^3$ or $10^4$ receptors per cell for an association rate of ~ $10^9 M^{-1} \min^{-1}$ or $10^8 M^{-1} \min^{-1}$, respectively) and we compared a homogeneous receptor distribution with a single cluster with ($\tilde{r}_0 = 17$) and multiple clusters ($\tilde{r}_0 = 5$) (Fig.4). We chose the clusters to be distributed randomly on the surface, but simulations with smaller lattices have shown that the dissociation curve is not significantly different for a regular, periodic arrangement of clusters also. In the real system, these clusters would have radii of approximately 25-90 nm respectively. Simulations were carried out with two levels of receptor density inside clusters: in the first case, rafts were occupied by receptors to saturation ($R'_0 = 1/\lambda^2$), and in the second case, the packing density was lowered to 0.1 ($R'_0 = 0.1/\lambda^2$). Clear differences, despite each system having the same actual density of receptors and surface coverage, are evident when clustering is present. In both the cases, there is clear evidence of a significant intermediate exponential regime (Figs 4A and 4B), which subsequently crosses over to a slower decay at later times. However, the effect of clustering on the dissociation rate is much more noticeable in the first case where the packing density of receptors is high (Fig 4C). Moreover, we see that for the high packing density case, the dependence of the effective rate (defined in the figure legend) on the cluster size observes the inverse linear relationship predicted by the theoretical analysis, Eq.19 (Fig. 4D).

The numerical results for the effective dissociation rates for the two cases discussed above may be put together in a single plot, by expressing the effective dissociation rate as a function of the ratio $\xi/\xi_0$. Clearly, for the same value of $\xi$ (~ raft radius), the threshold radius $\xi_0$ is different for the two cases (due to the inverse relationship to $R'_0$, Eq.17b) In fact by substituting the numerical values of the simulation parameters ($k_+ = D\lambda$), it is easily seen that $\xi_0 = 2\lambda$ for the case $R'_0 = 1/\lambda^2$ and $\xi_0 = 20\lambda$ for the case $R'_0 = 0.1/\lambda^2$. We may also use the equivalence with the Shoup-Szabo result (Eq.1) to express $\xi$ in terms of the cluster radius $\tilde{r}_0$: $\xi = (\pi/4)\tilde{r}_0$ from the previous discussion. In Fig 5, we plot the ratio of the effective dissociation rate, defined as the exponential fit to the initial straight portion of the data (t > 10), to the intrinsic rate $k_-^{eff}/k_-$ (after correcting for the lattice effects), which shows that this ratio is a smooth monotonically decreasing function of $\xi/\xi_0$. The theoretical prediction for the same is $1-\Sigma(0)$ (Eq.8), where $\Sigma(0)$ is given by Eq.17b, and is plotted as the smooth line in Fig.5. It is clear that the data points agree very well with our theoretical prediction in the regime $\xi/\xi_0 \geq 2$ (which is also the regime where clustering significantly alters the dissociation).

Fig.6 shows the effect of clustering in the high mean density case with $\theta = 0.1$ (~$10^5$ receptors per cell) and cluster radii of $\tilde{r}_0 = 10.0$ and $\tilde{r}_0 = 50.0$. A noticeable upward shift (decreased dissociation/increased binding retention) in the dissociation curve is observed, but the effect is non-monotonic and vanishes for small and large times, in both cases. This is illustrated more clearly in Fig.7 where we plot the difference between the bound fractions for clustered versus homogeneous receptor distributions as a function of time for the two values of the cluster radii. For the parameters used in the simulations ($\beta = k_-(\lambda^2/2D) = 10^{-4}, R_0 = 0.1/\lambda^2, k_+ = 0.1D\lambda$), the threshold cluster size is $\xi_0 \approx 20\lambda$ (i.e., $\tilde{r}_0 = 20$ in simulations) from Eq.17b. For $\tilde{r}_0 = 10.0$ and $\tilde{r}_0 = 50.0$ respectively, the parameter $\omega$ defined in Eq.24 takes values 0.9 and 4.5. For the first case (since $\omega < 1$), therefore, we also compared the simulation results with the approximate theoretical prediction in Eq.26 (smooth line in Fig.7), expected to be valid in the early-time regime. We observe that although the theoretical expression approximates the observed difference rather well at early times for small cluster size, it fails to capture the non-monotonous behavior at somewhat late times. It is likely that this dense mean receptor regime lies outside the applicability range of the expression in Eq.26. Clearly, a more systematic method to study the crossover from small to large receptor density would be desirable, but eludes us at this stage.

We now present a theoretical argument, which suggests that, over sufficiently long time scales, receptor clustering should have no effect on ligand dissociation, as found for the high density receptor case. Let us consider two different scenarios: (i) a homogeneous receptor distribution with a mean density $R_0$, and (ii) a clustered configuration, where the clusters have mean area density $Q_0 \approx R_0/n$, where $n$ is the average number of receptors per cluster. The first case was already studied in Sec.2, where we showed that the dissociation is characterized by a single time scale $c = \dfrac{D}{R_0^2}\left(\dfrac{k_-}{k_+}\right)^2$. Let us now map case (ii) into case (i), and imagine the clusters as effectively single receptors with mean density $Q_0$, and effective association and dissociation rates $k'_+$ and $k'_-$ respectively. The effective rates may be expressed in terms of the intrinsic rates using the Berg-Purcell-Shoup-Szabo relations, which give $k'_+ = nk_+(1-\gamma)$ and $k'_- = k_-(1-\gamma)$, where the `escape probability' $1-\gamma$ has been defined earlier (see Eq.1 and above). We now define the time constant for the clustered distribution as $c' = \dfrac{D}{Q_0^2}\left(\dfrac{k'_-}{k'_+}\right)^2$. Upon substituting for the primed quantities and the cluster density, we see that $c' = c$, i.e., the clusters have no effect on the decay at all! This analysis, however, is not exact and numerical simulations did show a significant effect of clustering in the strong rebinding case particularly at early times (insert, Figure 6). Thus, for the simple one-to-one ligand-receptor binding case it is conceivable that the effects of clustering are only transient but could still have a significant impact over a biologically relevant time scale.

## (ii) COMPARISON WITH EXPERIMENTS

Having compared the theoretical formulation in sufficient detail with lattice simulations, we turn to the question: How do the predictions of our simple model fit with experimental observations? We focus on the results of FGF-2 dissociation from HSPG obtained by Chu et al. (6), shown in Fig 1. FGF-2 binds to a high-affinity receptor FGFR as well as the HSPG we discuss here and higher order clusters including both FGFR and HSPG are possible (12). Therefore, any quantitative analysis of FGF-2 binding has to be done with care, because of the presence of competing interactions. In spite of this and because of a lack of availability of experimental dissociation data with other raft proteins, we choose this system for our analysis.

The experiments reported in (6) were done with intact cells either in the absence or presence of the lipid raft-disrupting agents MβCD and filipin (filipin data is not shown in Fig 1). Both lipid raft-disrupting agents were demonstrated to have a significant effect on the dissociation rate but we focus here on the MβCD data set since the mechanism of action is simpler and more straightforward. Briefly, a $k_-$ value of $0.004 \pm 0.002$ min**Error! Objects cannot be created from editing field codes.** was obtained for the control cells whereas treatment with MβCD increased the dissociation rate to approximately 0.023 min**Error! Objects cannot be created from editing field codes.**(with simple exponential fitting). If the MβCD treatment resulted in a completely homogeneous HSPG distribution, we arrive at a ratio of ~ 5.75 for the reduction in the dissociation rate due to raft-associated clustering.

The first question then is whether the present estimates of the HSPG surface density in these cells would allow for a significant exponential regime for the temporal decay of the dissociation curve? Using Eq.11b, we may compute the length of this time interval $t_e$ where the decay is exponential. Let us use the following estimates: $D \approx 10^{-11} - 10^{-10} m^2 s^{-1}$, $k_+ \sim 1.5 \times 10^6 M^{-1} s^{-1}$, $R_0 \sim 10^5 - 10^6 / l^2$, where $l \sim 5 \mu m$ is a rough estimate for the cell `radius'. After substitution in the expression in Eq.11a, these values give $t_e \approx 0.1 - 10 s$. This time scale is very small for typical dissociation measurements and suggests that the observed mode of decay in Fig 1 is more likely to be the non-exponential function predicted in Eq. 11b. More evidence for the presence of strong rebinding in the experiments shown in Fig 1 is seen when rebinding was prevented by the addition of heparin (Fig 1), which act as solution receptors for the released FGF-2. The dissociation in the presence of heparin was found to be increased compared to both untreated and MβCD treated and essentially the same with and without lipid raft disruptors (Fig. 1). Further, although limited, the data points suggest that dissociation could be exponential. To summarize, the difference between MβCD treated and untreated without heparin indicates an effect on dissociation by clustering and the heparin data suggests that rebinding is still an issue even in the absence of rafts.

It is important to note that because of the slow, non-exponential decay of the dissociation curve in the presence of strong rebinding, this function cannot be accurately characterized by a single rate valid over a well-defined time regime (unlike the weak-rebinding case). Rather, the effective rates obtained by fitting the experimental curves to exponential functions are only a simplified characterization of the decay valid over a limited time scale. Keeping these caveats in mind, we tried to see whether the observed experimental data, with and without raft disrupters, is reproduced by the theoretical functions of Eq.11b (homogeneous distribution) and Eq.26 (raft-correction). The curves that were judged to be closest to the experimental data in Fig 1 (by comparing with the exponential fit functions used to estimate the dissociation rates in Fig 1) are shown in Fig. 8. The parameters $c$ and $k_-\omega$ (Eq.11b and Eq.26) were tuned for the best fit, and the optimal numerical values found were $c = 1.1 \times 10^{-4} s^{-1}$ and $k_-\omega = 4 \times 10^{-4} s^{-1}$. Let us now substitute for the following parameters: $D = 10^{-11} m^2 s^{-1}$, $k_- = 0.25 s^{-1}$ (obtained from the heparin data in Fig 1), $k_+ = 1.5 \times 10^6 M^{-1} s^{-1}$ (11). We treat the surface densities $R_0 \approx N/l^2$ (where $l \approx 5 \times 10^{-6} m$ is the typical cellular dimension) and $R'_0 \approx 10/\xi^2$ as unknowns, where $N$ is the total number of HSPG per cell and $\xi$ is roughly the radius of a raft. Upon solving for the unknowns $N$ and $\xi$, we find $N \approx 7.5 \times 10^5$ and $\xi \approx 200 nm$. Both values are within reasonable limits of the known estimates of these parameters, and the resemblance between Fig 8 and Fig 1 supports the FGF2-HSPG system analysis under the strong rebinding category discussed in Sec 2, case 2. The implications of this observation are (i) the effective dissociation rate measured in experiments with such high coverage receptors such as HSPG is best treated as a phenomenological parameter valid for a limited time range, (ii) the theory can be used with the experimental observations in order to determine the true dissociation rate, and (iii) the signaling events where rafts are expected to play a role may be expected to occur over time scales where the transient effects of clustering are still relevant.

Suppose however, as an aside, that Sec 2, case 1 (low surface coverage) would have applied to this experimental system. From Fig 5, we note that a reduction in the effective dissociation rate by a factor ~5.75 (or a ratio of 0.17) for a low density system would require that the ratio $\xi/\xi_0$ should be around 2.87. Let us now use Eq.17b to express this result in terms of the raft radius $r$ by means of the substitutions $r = (4/\pi)\xi$ and $R'_0 = n/\pi r^2$ where $n \sim 5-10$ is a rough estimate of the number of HSPG per raft. The condition that $\xi/\xi_0 \approx 2.87$ now demands that (for $r \sim 25 nm$) the association rate for FGF2-HSPG should be nearly $k_+ \sim 3.44 \times 10^8 (10^9) M^{-1} s^{-1}$ for $D \sim 10^{-10}(10^{-11}) m^2 s^{-1}$ respectively (we allow some flexibility in D, because by definition, D is actually 1/3 of the real three-dimensional diffusion coefficient). This value is an order of magnitude or two larger than the available experimental number for HSPG: $k_+^{exp} \sim 1.5 \times 10^6 M^{-1} s^{-1}$ (10). However, it must be noted that although the above theoretical estimate is somewhat high for FGF-2-HSPG, it is still within the range of association rates typically reported in the literature. We believe that, therefore, there could well be other low-density raft proteins that could use the enhanced-rebinding mechanism in order to retain ligands longer near the surface and for which our theory could be useful.

Let us now address the following question: Is there likely to be a long-term effect of rafts on ligand dissociation for FGF-2-HSPG based on the analysis of the system? The numerical simulations coupled with the theoretical argument presented in the previous section showed that the effect of clustering for our model system was present only in a limited time range and vanished at late times. Experiments however did not support this for the FGF-2-HSPG-lipid raft system. This system however is much more complex than the model system our theoretical and numerical analysis were based on primarily due to the multiplicity of receptors (i.e., FGF receptors and HSPG competing for FGF-2 binding as well as forming higher order complexes). That being said, our systematic study of diffusion-controlled rebinding in the presence of receptor clusters indicates the limitations of this mechanism: the surface coverage or the association rate of the receptors have to be sufficiently large in order to have a measurable impact of clustering. It is therefore worthwhile to explore alternative mechanisms that might be employed by the cell to increase ligand retention inside rafts. In the last part of this section, we will discuss one such plausible mechanism whereby the ligands may be retained longer inside a cluster, i.e., internal diffusion of ligands inside a cluster of receptors. We emphasize that the model is a theoretical idea and not strictly based on experimental observations.

### (iii) INTERNAL DIFFUSION MODEL

An alternative model for ligand dissociation in the presence of clusters is now proposed, by invoking a 'co-operative rebinding' mechanism for ligand retention inside a cluster. For example, FGF-2 and other heparin-binding growth factors bind HSPG through the glycosaminoglycan (GAG) side-chains. In this model, we would propose that there might be overlap of the GAG chains on neighboring HSPG clustered in rafts, resulting in a preferential path whereby a ligand, following its release from one GAG binding site, might find it energetically more favorable to bind to a neighboring binding site belonging to another HSPG. The ligand would therefore perform a surface diffusion inside the cluster, and likely be released into the solution only upon reaching the edge of the cluster. Clearly, this 'internal diffusion' would significantly reduce the effective dissociation rate of the ligand, as we show now more quantitatively.

For simplicity, let us imagine the binding site inside a raft as occupying the sites of a lattice with spacing $d$, which is the typical separation between two molecules. A cluster of radius $r$ has $n \sim (r/d)^2$ molecules inside it. Let us now assume that the 'hopping' of the ligand from one site to another takes place over a mean time interval $\tau$. Then, the diffusion coefficient for the surface diffusion of the ligand inside the cluster is $D_s \approx d^2/\tau$. The total time it takes the ligand to reach the edge of the cluster by internal diffusion is, therefore,

$$T \approx r^2/D_s \sim n\tau \ . \tag{27}$$

The ligand is likely to fully dissociate from the cluster once it reaches the edge, since there is less likelihood of finding a neighboring site to bind to. Thus, the mean effective dissociation rate is given by

$$k_-^{eff} \sim T^{-1} \sim v/n \ , \tag{28}$$

where we have also defined the internal hopping rate $v \sim \tau^{-1}$.

Although it is difficult to have an independent estimate for $v$, it appears reasonable to assume that this is of the same order as the intrinsic dissociation rate $k_-$ for individual receptors. In this case, if the number of HSPG per cluster is $n$, then the dissociation rate is roughly reduced by a factor of $1/n$, which could then account for the experimentally observed ratio of ~1/6.

In Fig.9A, we show some numerical simulation results done with this 'internal diffusion' model. These simulations were done with a mean surface coverage of $\theta = 0.001$. The main figure shows the comparison between the dissociation curves obtained with the rebinding model and the internal diffusion model for cluster radius $\tilde{r}_0 = 5$, whereas the inset shows the same for $\tilde{r}_0 = 8$. The figures show a much more dramatic effect of clustering on dissociation as compared to the purely diffusion-limited rebinding model which has been the main subject of this paper. For instance, for $\tilde{r}_0 = 5$, the rebinding model results in a reduction in the effective dissociation rate by a factor of ~ 0.21, whereas in the internal diffusion model, the corresponding number is ~ 0.0019, i.e., lower by two orders of magnitude. Similar trends were seen for other values of the cluster radii also. Fig. 9B shows the effective dissociation rate (found by fitting the data in Fig.8A to exponential curves) in the model plotted as a function of the number of proteins `n' inside a cluster. In accordance with our arguments, we see a sharp drop of the decay rate with `n', but the curve is non-linear and does not fit completely with the simple 1/n dependence predicted in Eq.28. Nevertheless, it is obvious that such co-operative mechanisms could greatly augment the effect of receptor clustering, and we speculate that lipid rafts possibly use a combination of both enhanced rebinding as well as more specific ligand retention mechanisms to slow down the dissociation.

Although there is no direct experimental evidence for any effective `confinement' of FGF-2 within the HSPG clusters, it is possible that such additional mechanisms could be present in this or other systems to enhance the purely diffusion-controlled rebinding described earlier. Models of surface diffusion of ligands on receptor clusters have been discussed in the literature in other contexts as well, e.g., `molecular brachiation' of CheR molecules on a cluster of its receptor proteins (35). A later model of ligand association to a cell surface containing receptor clusters, incorporating such an internal diffusion mechanism (26) had been found to explain experimental data (35) better than previous models (18, 19) that did not explicitly contain such mechanisms.

Finally, what could be the possible advantage of the internal diffusion mechanism, over simple enhanced rebinding due to clustering? We believe that it is primarily the effect on time scales. The increased rebinding due to clustering leads to a significant effect on the effective dissociation rate, but only over certain limited time scales, as we showed in detail in the preceding section. By contrast, the internal diffusion/trapping mechanism could cause a permanent reduction in the dissociation rate, as long as clusters are present. The limitation, of course, is that the receptors have to be packed rather tightly inside a raft for this mechanism to take effect.

## (IV) DISCUSSION

It is generally understood that lipid rafts are capable of confining several kinds of large proteins inside them for time scales up to several minutes (2). HSPG are among the proteins shown to localize to lipid rafts (and they are also co-receptors for heparin-binding growth factors such as FGF-2). We therefore sought to determine theoretically whether the confinement and clustering of HSPG inside lipid rafts could affect binding of FGF-2, either via promoting rebinding of dissociated ligands and/or via reduced dissociation through some co-operative interactions between HSPG in close proximity to each other. Work by Chu et al. (2004) indicated that lipid rafts play a significant role in controlling the dissociation of FGF-2 from HSPG, but the mechanism behind this effect is speculative (6).

In this paper, we present a rigorous mathematical formalism to study the rebinding of ligands to receptors on an infinite plane, as an approximation of the surface of a tissue culture plated cell. In contrast to work by Lagerholm and Thompson (27) who employed partial differential equations to describe the time evolution of the space-averaged ligand density, we have adopted a stochastic formalism, and described the dynamics in terms of the return-to-the-origin characteristics of the Brownian trajectories of the ligands. However, the theory is constructed entirely in terms of coarse-grained continuum variables, which constitutes an improvement over our previous model (28), which was based on a lattice random walk. We predict that the long-time decay of the bound fraction always assumes a non-exponential form for the planar surface studied here, irrespective of any parameter values. However, the entry into this regime depends on the association rate and the surface density of receptors. The theory also recovers the existence of an exponential regime at early times. We have checked and confirmed these analytical results through numerical simulations.

The principal aim of this paper was to utilize this formalism to study the effect of large-scale receptor clustering on the cell surface, as appears to occur, for example, inside lipid rafts. We have quantified the reduction in the effective dissociation rate due to such clustering in various cases of interest. In the regime of low mean receptor density, our predictions agree with earlier results obtained by means of different arguments (19). Monte Carlo simulations provide excellent support for our model. A direct comparison with experimental results for the high mean receptor density case was also done noting that there is a lack of experimental data currently available in the literature for systems which might better be described by our theoretical model (i.e. small monovalent ligand interacting with a single transmembrane receptor which does not dimerized or form higher order complexes). With further refinements, our theory could provide an independent method to check for spatial non-uniformity in receptor distribution on the cellular surface. This is intimately related to the much larger question of cell membrane organization, a subject of much debate and discussion in recent times (3, 37, 38). Even so, as a first attempt to explicitly study the impact of enhanced rebinding due to large-scale assembly of receptors on the dissociation rate of ligands, we believe that our findings are of value to both experimentalists and modelers interested in lipid rafts and their role in cell signaling.

Our theoretical formalism was developed for the case of a planar substrate, and most of our results are specific to this geometry (with the exception of the section 'Extension to receptor clusters: Case 1', which also applies to other geometries, eg. spherical cells, provided the mean surface density of receptors is small). Apart from the obvious suitability of this geometry to many experimental situations, the calculations could be effectively reduced to one dimension, which greatly simplified the analysis. It would be interesting, albeit challenging (on account of the angular dependence of the probabilities), to extend the theory presented in this paper to the case of receptors on a spherical cell surface. Numerical studies in this direction are currently being carried out. A better characterization of the different crossover regimes in the present theory is also desirable.

We have assumed that receptors are "stationary" and do not exit clustered zones. How stable is the association of a protein to the raft? Single-particle tracking experiments have shown that the diffusion of a raft-associated protein is unchanged over times scales of up to 10 min, indicating that the proteins can remain with the raft during this period (32). However, since the dissociation measurements typically extend considerably longer (on the order of hours), the possibility of the proteoglycans exiting the raft during this period cannot be ruled out. It would be interesting to see, in a future study, if such dissociation events could have any impact on the rebinding process by rendering the surface coverage factor time-dependent inside rafts. Other relevant issues that would be worthwhile investigating in this context include the effects of raft diffusion and their stability. It is also straightforward to pose questions about noise within our formalism. For example, characterization of fluctuations of receptor occupancy (including temporal correlations) would be relatively easy to address in our model (being based on individual ligand histories) and could provide insight into the much broader question of how well does the cell sense its environment.

In conclusion, we have presented a novel theoretical framework to study the problem of ligand rebinding to receptors on the cellular surface, and how the rebinding and effective dissociation of ligands are regulated by the spatial organization of receptor proteins. Although many specific results in this paper are restricted to ligand binding to a planar surface, the framework itself is more general, and could be generalized to other cases, e.g., isolated spherical cells which are more suitable for some situations (e.g., the immune system).

We would like to acknowledge H.J. Hilhorst, Y. Kafri, R. Kree, D. Lubensky, M. Rao, B. Schmittmann, and M. Zapotocky for fruitful discussions. Early assistance with simulations from Eric Spiegel and Satheesh Angaiah is appreciated. Financial and computational support from the Max Planck Society (MG), National Science Foundation [awards NSF-DMR 0089451 (MG),

NSF-9875626 (KFW), and NSF-DMR 0308548 (UCT)], National Institutes of Health [NIH-HL56200 (MAN)], and from the Bank of America Jeffress Memorial Trust [Grant no. J-594 (UCT)] is gratefully acknowledged.

## Appendix A

In this appendix, we briefly outline the calculation of the probability density of first returns to the origin. As is conventional, we consider a one-dimensional random walk starting from the origin at time $t = 0$. Let $Q(t)$ be the probability density at the origin at time $t$, and let furthermore $q(t)$ denote the density of walkers that return to the origin for the first time at time $t$.

As in the section 'Rebinding on a planar surface', we compartmentalize the available space into cells of volume $\lambda^3$ and approximate the continuum diffusion as hopping between adjacent cells. Then, the first passage probability itself is simply $\lambda q(t)$, and the two probability densities are therefore related in the following way.

$$2\lambda Q(t) = \lambda q(t) + \frac{\lambda}{2\delta} \int_{2\delta}^{t} d\tau\, q(\tau) Q(t-\tau) 2\lambda \quad . \tag{A1}$$

The $\lambda$ factors are written explicitly for the sake of clarity. The difference of 2 in the measure of length arises because the first passage events (by definition) cannot cross the origin.

In terms of Laplace transforms (in the limit $\delta \to 0$), the relation becomes

$$\tilde{q}(s) = \frac{2\tilde{Q}(s)}{1 + \frac{\lambda}{\delta}\tilde{Q}(s)} \quad . \tag{A2}$$

Given that $Q(t) = (4\pi Dt)^{-1/2}$, $\tilde{Q}(s) = \frac{1}{\sqrt{4Ds}}$, and it follows that

$$\tilde{q}(s) = \frac{1}{\sqrt{Ds} + \frac{\lambda}{2\delta}} \quad . \tag{A3}$$

The explicit inversion of this transform gives (28)

$$q(t) = \frac{1}{\sqrt{\pi Dt}} - \frac{a}{\sqrt{D}} e^{a^2 t} \operatorname{erfc}(a\sqrt{t}) \tag{A4}$$

where $a = \lambda/(2\delta\sqrt{D})$. The late time behavior of this quantity is given by (29)

$$q(t) \approx \frac{2D}{t}\left(\frac{\delta}{\lambda}\right)^2 \frac{1}{\sqrt{\pi Dt}} \quad ; \quad t \gg \delta \quad . \tag{A5}$$

The $t^{-3/2}$ behavior is consistent with the well-known result for the first passage probability in the context of one-dimensional random walks (see, e.g., 39).

## Appendix B

In this appendix, we explore the temporal behavior of $C_{R_0}(t)$. From Eqs.8 and 9, we infer its Laplace transform to be $\tilde{C}(s) = [\sqrt{Ds} + k_+ R_0]^{-1}$. The explicit inversion reads (28)

$$C_{R_0}(t) = (\pi D t)^{-1/2} - \frac{k_+ R_0}{D} \exp[(k_+ R_0)^2 t / D] \, \text{erfc}[k_+ R_0 \sqrt{t/D}] \, , \tag{B1}$$

which has the following limiting forms (29):

$$C_{R_0}(t) = (\pi D t)^{-1/2} \, , \qquad \text{when} \quad t \ll \frac{D}{(k_+ R_0)^2} \, , \tag{B2a}$$

$$C_{R_0}(t) \approx \sqrt{\frac{D}{4\pi}} (k_+ R_0)^{-2} t^{-3/2} \, , \quad \text{when} \quad t \gg \frac{D}{(k_+ R_0)^2} \, . \tag{B2b}$$

Clearly, two distinct time regimes may be identified here. When $k_+ R_0$ is small, absorption by the surface becomes rare, and the first term dominates in the expression at sufficiently small times. In this case, the probability density $C_{R_0}(t)$ is the same as for a perfectly reflecting surface (39). In the converse limit, absorption is dominant and the temporal behavior exhibits the $t^{-3/2}$ dependence characteristic of the probability density for a perfectly absorbing surface (39).

## Appendix C

In this appendix, we estimate the effect of rebinding on ligand dissociation from a single isolated receptor in an infinite cubic lattice. The ligand dissociates from the receptor with probability $\beta$ and performs a random walk on the lattice, until the walk hits the (stationary) receptor again and binds to it. We are interested in estimating the probability $p(N)$ that the ligand is bound to the receptor after $N$ time steps. (One time step is the time required for the ligand to move one lattice spacing.)

The general equation for $p(N)$ is

$$p(N+1) = p(N)[1-\beta] + \beta \sum_{2}^{N-2} p(k) C(N-k) \, , \tag{C1}$$

where $C(k)$ is the probability of return to origin of a three-dimensional random walk (Polya walk) after $k$ time steps.

Let us now make the reasonable assumption that the function $p(N)$ is monotonically decreasing with $N$, in which case $p(k) \geq p(N)$ in Eq.C1. This means that

$$p(N+1) - [1-\beta] p(N) \geq \beta p(N) \sum_{2}^{N-2} C(k) \, . \tag{C2}$$

We next consider the limit of large $N$, in which case the sum in Eq.C2 becomes the probability that the random walk will *ever* return to the origin, which is nearly 0.3403 (39). In this limit, we may also treat $N$ as a continuous variable, and make use of the approximate replacement $p(N+1) - p(N) \approx dp(N)/dN$ which gives

$$\frac{dp}{dN} \geq -0.66\beta p(N) \quad , \tag{C3}$$

which means that

$$p(N) \geq p(N_0) exp[-0.66\beta(N-N_0)] \text{ for } N, N_0 \gg 1 \quad . \tag{C4}$$

If the late-time behavior is characterized by an effective exponent $\beta_{eff}$, Eq.C4 shows that this exponent is bounded by the relation

$$\beta_{eff} \leq 0.66\beta \quad . \tag{C5}$$

| Quantity | Symbol | typical units |
|---|---|---|
| Microscopic length scale | $\lambda$ | m |
| Diffusion coefficient | D | $m^2 s^{-1}$ |
| Microscopic diffusion time scale | $\delta t = \lambda^2 / 4D$ | s |
| Association rate | $k_+$ | $M^{-1} s^{-1}$ |
| Dissociation rate | $k_-$ | $s^{-1}$ |
| Equilibrium dissociation constant | $K_D = \dfrac{k_-}{k_+}$ | M |
| Mean surface density of receptors | $R_0$ | Number of molecules/$m^2$ |
| Surface density of receptors Inside clusters | $R'_0$ | Number of molecules/$m^2$ |
| Bound receptor fraction at time t | $p(t)$ | dimensionless |
| Ligand density profile close to the surface at time t | $\rho(t)$ | Number of molecules/$m^3$ |
| Return to origin probability density for a surface with $R_0$ receptors per unit area | $C_{R_0}(t)$ | $m^{-1}$ |
| Return to origin probability density for a perfectly absorbing surface | $q(t)$ | $m^{-1}$ |
| Probability of non-absorption upon contact | $\gamma$ | dimensionless |
| Time scale of exponential decay (ref: Eq.11a) | $t_e$ | s |

**TABLE 1**: A glossary of the important quantities discussed in the paper, along with the corresponding units (m=meter, s=seconds, M=mole)

|  | Quantity | Dimensionless form |
|---|---|---|
| Surface density | $R_0$ | $\theta = R_0 \lambda^2$ |
| Association Rate | $k_+$ | $\tilde{k}_+ = 2k_+ \delta t / \lambda^2$ |
| Dissociation Rate | $k_-$ | $\tilde{k}_- = k_- \delta t$ |
| Cluster size | $r_0$ | $\tilde{r}_0 = r_0 / \lambda$ |
| Diffusion coefficient | $D$ | $\tilde{D} = D \delta t / \lambda^2 = 1$ |

**TABLE 2**. A list of the dimensionless forms of various quantities, scaled using the length scale $\lambda$ and time scale $\delta t = \lambda^2 / 2D$, respectively. Typical numerical values are $\lambda \approx 1-5\,\text{nm}$, $D \sim 10^{-6}\,cm^2 s^{-1}$, $k_+ \sim 10^5 - 10^8 M^{-1} s^{-1}$, $k_- \sim 1 - 10^{-4}\,s^{-1}$, and $r_0 < 100\,\text{nm}$ (estimates for lipid rafts, reviewed in 4).

**FIGURES**

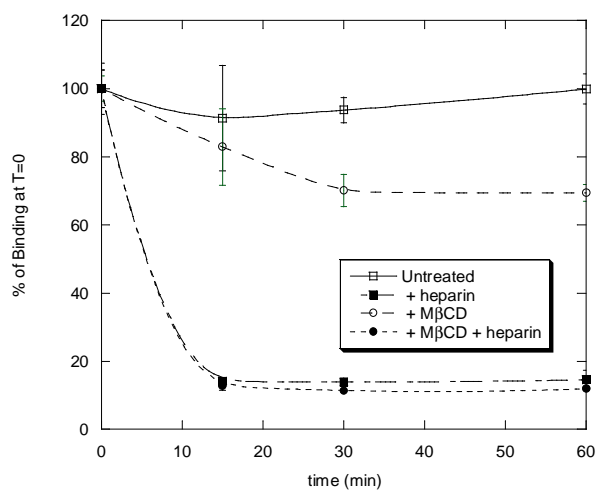

**FIG 1**. Effect of the Lipid Raft Disrupting Agent MβCD and heparin on FGF-2 dissociation from HSPGs. Bovine vascular smooth muscle cells in tissue culture were treated with MβCD (0 (untreated) or 10 mM (MβCD)) for 2h at 37°C prior to cooling to 4°C. $^{125}$I-FGF-2 (0.28 nM) was added and allowed to bind to the cells for 2.5 hr prior to initiation of dissociation (t=0). After the binding period, unbound $^{125}$I-FGF-2 was removed by washing the cells with cold binding buffer, and dissociation was initiated in binding buffer without FGF-2 (± heparin (100 μg/ml)) at 4°C. The cells were allowed to incubate for the indicated time periods at which point the amount of FGF-2 bound to HSPG sites was determined by extracting the cells with 2M NaCl, 20 mM Hepes, pH 7.4 and counting the samples in a gamma counter. All data was normalized to the amount of $^{125}$I-FGF-2 bound to HSPG sites at t=0 (100%) under each condition. Mean values of triplicate samples ± SEM are shown (data re-plotted from 5).

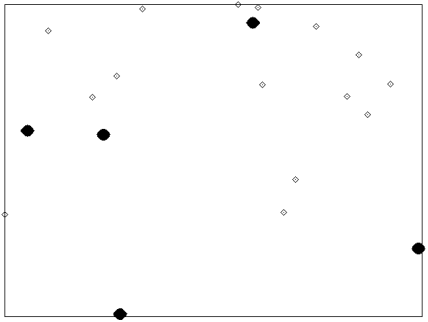 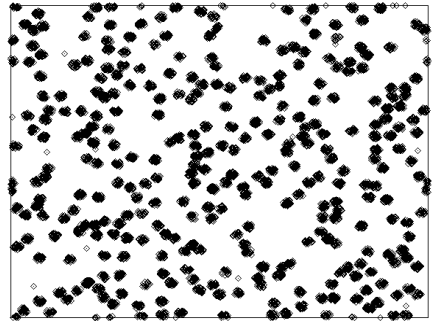

**A**.    **B**.

**FIG 2.** Two typical receptor configurations used in the Monte Carlo simulations. The mean receptor density (in dimensionless units) in A is 0.001 and in B it is 0.1. The cluster radius is $\tilde{r}_0 = 8.0$ in A and $\tilde{r}_0 = 10.0$ in B. The small dots are single receptors. The clusters are filled to saturation in both the cases.

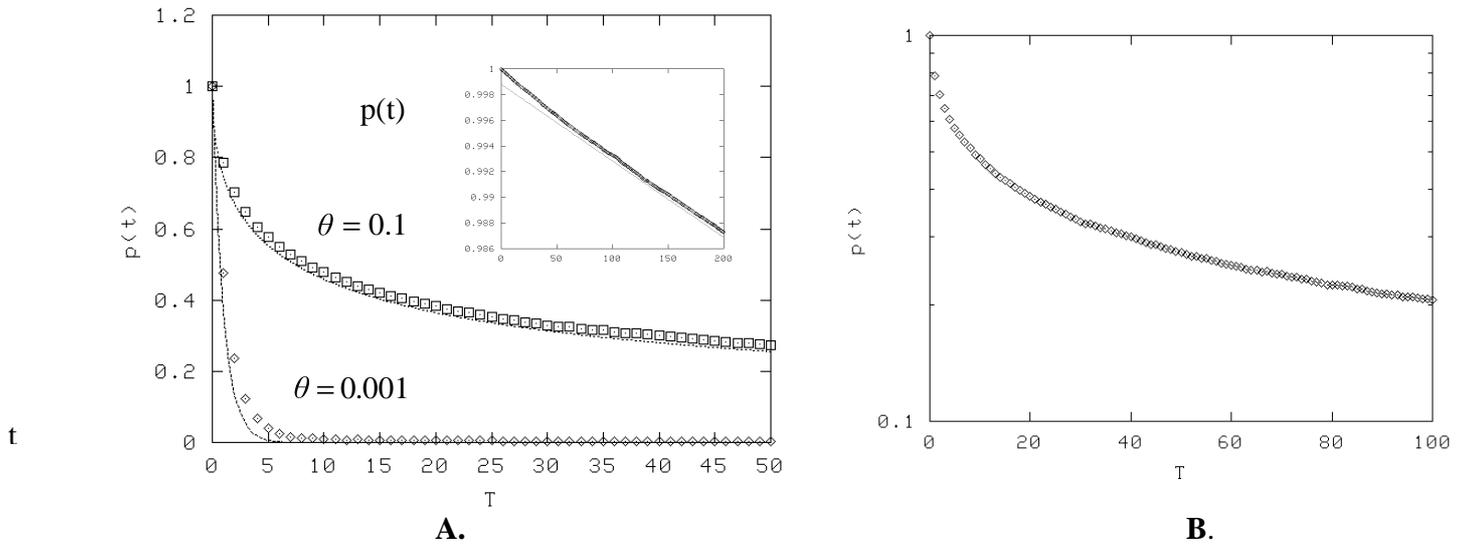

**FIG3**. (A). Receptor density impacts ligand dissociation for a uniform distribution of receptors. p(t) versus scaled time $T = \beta t$ (t is measured in number of Monte Carlo time steps) is plotted for $\theta = 0.001$ and $\theta = 0.1$ The former displays exponential decay, while the latter is clearly non-exponential over the time scales shown. The lines are theoretical fits: $e^{-k_- t}$ in the former case and the function in Eq.11b in the latter case with c=0.08 (the theoretical value from Eq.11 is 0.06). The early time behavior of the high-density case ($\theta = 0.1$) plotted in the inset figure does indicate exponential decay (inset: t is the number of Monte Carlo steps), but the effective dissociation constant is about $0.6 k_-$, less than the theoretical value $k_-$, see Eq.11a, also Appendix C. (B) The high-density ($\theta = 0.1$) data plotted on a semi-logarithmic scale, which shows more explicitly the strongly non-exponential nature of the decay.

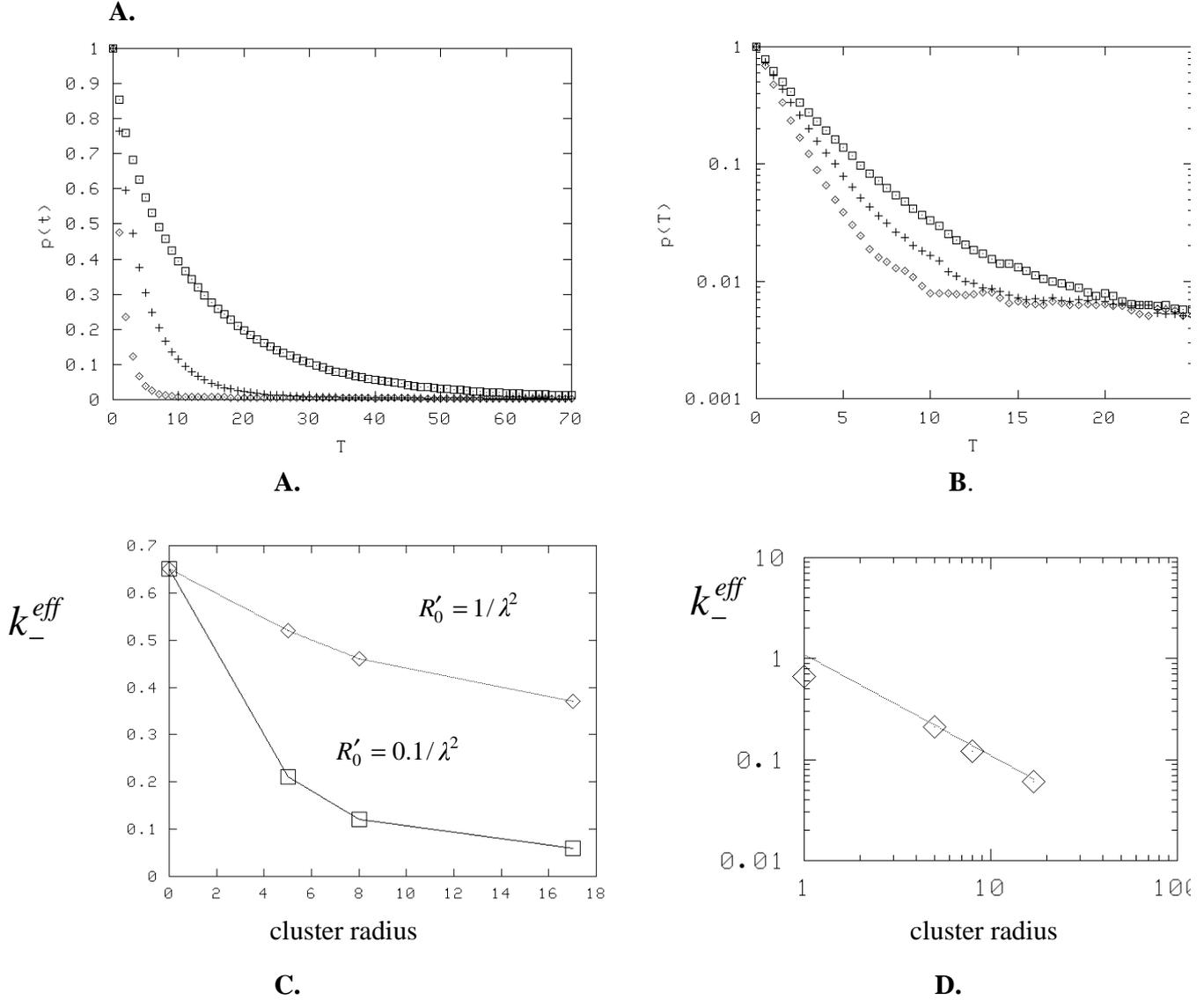

**FIG 4**. Dissociation is impacted by the degree of clustering when there is low mean surface density ($\theta = 0.001$). (A) Shown is p(t) versus scaled time where the curves correspond to uniform distribution, $\tilde{r}_0 = 5.0$, and $\tilde{r}_0 = 17.0$ (a single cluster in the last case), when the clusters are packed to saturation. The decay is exponential except for very late times, as is evident from the data plotted in semi-logarithmic scale in the inset. The axis labels are common to the main figure and the inset. (B) Similar data as in (A), but the packing density inside clusters is only 0.1, on semi-logarithmic scale. (C) Effective decay constant (exponential fit to the early portion (straight part) of the data) as a function of cluster radius for cases (A) and (B). (D) Effective decay constant for (A) plotted against cluster radius on a logarithmic scale. The straight line is a fit function proportional to $\tilde{r}_0^{-1}$, and the good agreement supports Eq.19. The slope for the uniform case ($\tilde{r}_0 = 1.0$) in (A) and (B) is ~ 0.67, which is less than the theoretical value 1, presumably due to (unavoidable) lattice effects in the simulations (for details, see Appendix C).

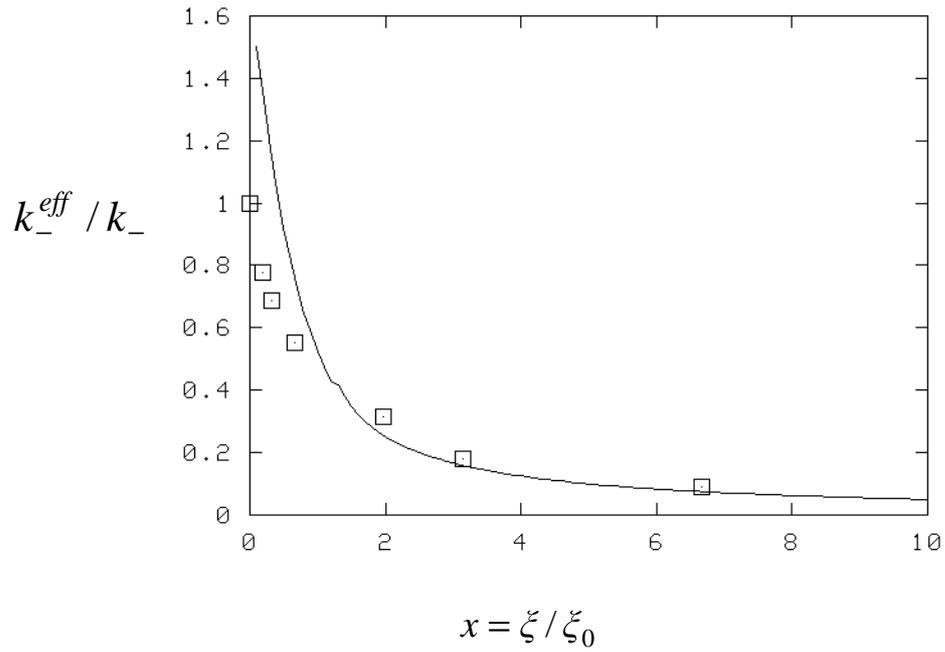

**FIG 5.** The effective decay constants (defined in the legend of Fig.4) plotted in 4C is plotted against the scaled cluster radius $x = \xi/\xi_0$. The smooth line is the theoretical fit function $1 - \Sigma(0)$, as defined by Eq.17a.

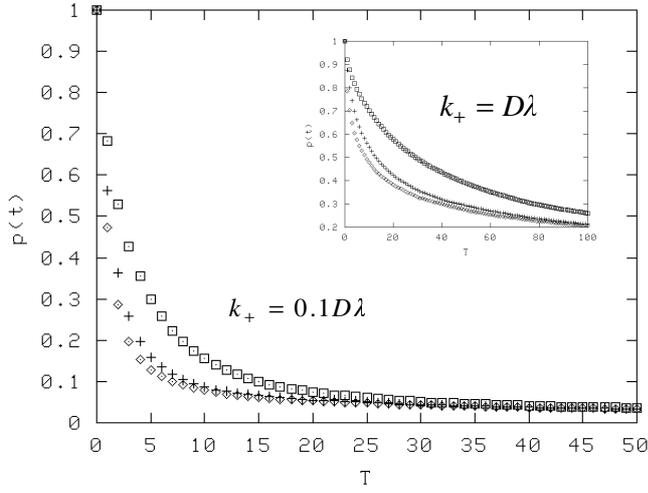
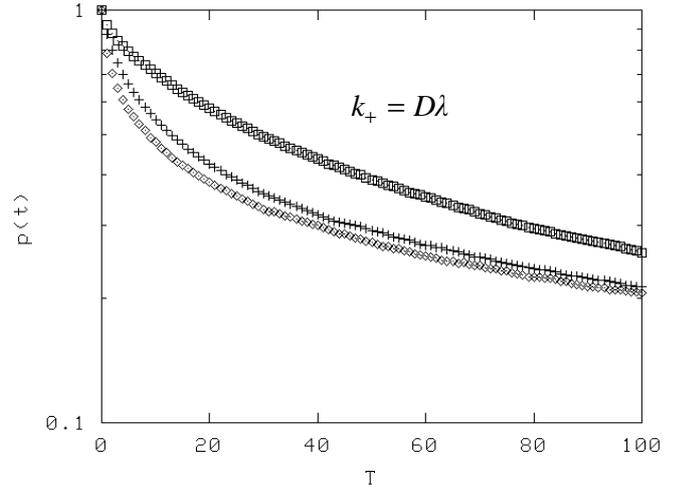

**A**.  **B**.

**FIG. 6**. . (A). Effect of clustering on the dissociation rate for the high mean surface density case ($\theta = 0.1$). p(t) versus scaled time is shown for two values of association rates: $k_+ = 0.1 D\lambda$ (main figure) and $k_+ = D\lambda$ (inset) for uniform distribution ( ), $\tilde{r}_0 = 10.0$ (+), and $\tilde{r}_0 = 50.0$ (□). The lower association rate in the main figure was used to increase the threshold cluster size (ref: Eq.17b) in order to verify the theoretical predictions in the section 'Extension to receptor clusters: Case (ii)'. The axis labels are common to the main figure and the inset.(B) The data in the inset of (A) is plotted on a semi-logarithmic scale to show the non-exponential nature of the decay more explicitly.

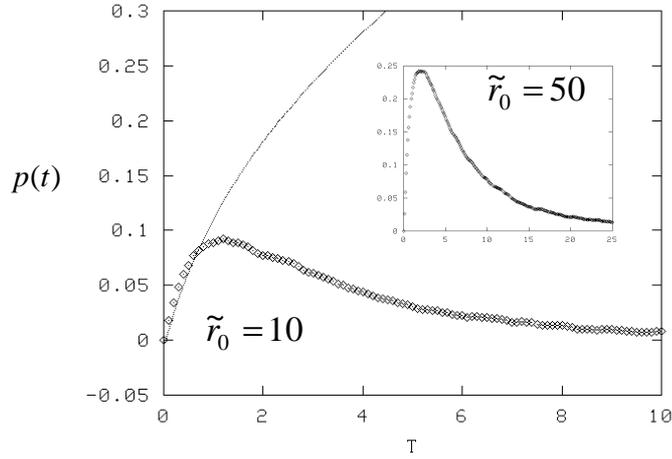

**FIG 7**. The difference $p(t)$ in the bound fraction between uniform distribution and clustered configurations when $k_+ = 0.1D\lambda$, corresponding to $\tilde{r}_0 = 10.0$ (main figure) and $\tilde{r}_0 = 50.0$ (inset) is plotted against the scaled time $T = k_- t$, along with the theoretical prediction from Eq.26 (thin line in the main figure). The theoretical curve agrees with the simulations in the very early regime, but deviates at later times. The impact of clusters vanishes at late times, in accordance with our arguments in the section 'Extension to receptor clusters: Case (ii)'. The axis labels are common to the main figure and the inset.

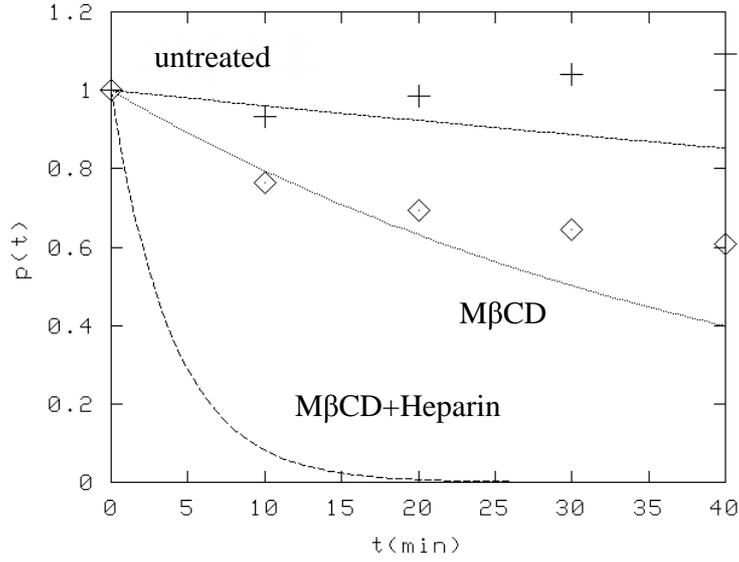

**FIG 8**. The three smooth curves in the figure represent the best exponential fits to the experimental data from Fig 1, with time constants (in units of 1/s) 0.25 (MβCD+Heparin), 0.023 (MβCD) and 0.004 (untreated). The data points represent the best theoretical fits, using the function in Eq.11b for MβCD, and with the added raft correction (Eq.26) for untreated cells. The fit parameters are $c = 1.1 \times 10^{-4} s^{-1}$ in Eq.11b and $\omega = 1.6 \times 10^{-3}$ in Eq.26. The corresponding values of the intrinsic variables are discussed in the main text, in Sec.III.(ii). Note that, at late times, some of the data points for the clustered configuration cross 1, which indicates that the limit of applicability of the perturbation theory has been reached, and that higher order terms in the perturbation series have to be taken into account.

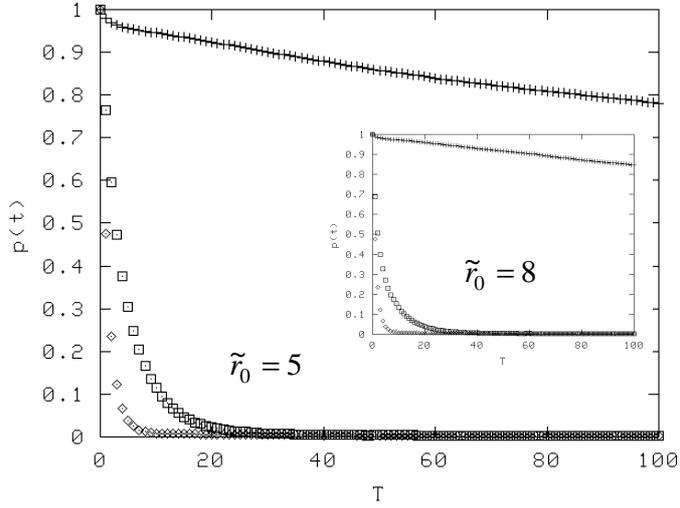

A.

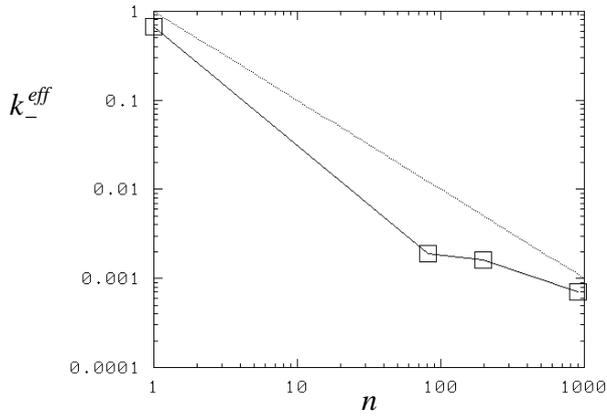

B.

**FIG. 9.** Clustering has a more significant effect on dissociation with 'internal diffusion' model. (A). p(t) versus the scaled time comparing the uniform distribution ( ) with that of clustered configurations for the internal diffusion model (+) and the rebinding model (□) for clusters of radius $\tilde{r}_0 = 5.0$ (main figure) and $\tilde{r}_0 = 8.0$ insert). (B). The effective decay constant for various cluster sizes using the internal diffusion model plotted on a log-scale as a function of the number of receptors in each cluster for the internal diffusion model. The straight line has slope -1, and is meant for comparison with the theoretical argument in Eq.28.

39